\magnification=1200
\baselineskip=13pt
\overfullrule=0pt
\tolerance=100000
\nopagenumbers

\font\tenbifull=cmmib10 \skewchar\tenbifull='177
\font\tenbimed=cmmib7   \skewchar\tenbimed='177
\font\tenbismall=cmmib5  \skewchar\tenbismall='177
\textfont9=\tenbifull
\scriptfont9=\tenbimed
\scriptscriptfont9=\tenbismall

\mathchardef\alpha="710B
\mathchardef\beta="710C
 \mathchardef\gamma="710D
\mathchardef\delta="710E
\mathchardef\epsilon="710F
\mathchardef\zeta="7110
\mathchardef\eta="7111
\mathchardef\theta="7112
\mathchardef\iota="7113
\mathchardef\kappa="7114
\mathchardef\lambda="7115
\mathchardef\mu="7116
\mathchardef\nu="7117
\mathchardef\micron="716F
\mathchardef\xi="7118
\mathchardef\pi="7119
\mathchardef\rho="711A
\mathchardef\sigma="711B
\mathchardef\tau="711C
\mathchardef\upsilon="711D
\mathchardef\phi="711E
\mathchardef\chi="711F
\mathchardef\psi="7120
\mathchardef\omega="7121
\mathchardef\varepsilon="7122
\mathchardef\vartheta="7123
\mathchardef\varphi="7124
\mathchardef\varrho="7125
\mathchardef\varsigma="7126
\mathchardef\varpi="7127

 at 8truept

\

\centerline{}
\bigskip
\bigskip
\bigskip
\baselineskip=18pt

\centerline{\bf On the Integrability of a Class of Monge-Amp\`ere Equations}
\vfill
{\baselineskip=11pt
\centerline{J. C. Brunelli\footnote{*}{brunelli@fsc.ufsc.br}}
\medskip
\medskip
\centerline{Universidade Federal de Santa Catarina}
\centerline{Departamento de F\'\i sica -- CFM}
\centerline{Campus Universit\'ario -- Trindade}
\centerline{C.P. 476, CEP 88040-900}
\centerline{Florian\'opolis, SC -- BRAZIL}
\bigskip\bigskip
\centerline{M. G\"urses \footnote{**} {gurses@fen.bilkent.edu.tr} and K. Zheltukhin \footnote{***}
{zhelt@fen.bilkent.edu.tr}}
\medskip
\centerline{Department of Mathematics,}
\centerline{Bilkent University,}
\centerline{06533, Ankara, Turkey}
}
\vfill

\centerline{\bf {Abstract}}

\medskip

We give the Lax representations for for the elliptic, hyperbolic and
homogeneous second order Monge-Amp\`ere equations. The connection
between these equations and the equations of hydrodynamical type give
us a scalar dispersionless Lax representation. A matrix dispersive Lax
representation follows from the correspondence between sigma models, a
two parameter equation for minimal surfaces
and Monge-Amp\`ere equations. Local as well nonlocal conserved
densities are obtained.
\medskip

\vfill
\eject
\headline={\hfill\folio}
\pageno=1
\bigskip
\noindent {\bf 1. {Introduction}}
\medskip

The nonlinear partial differential equation in $1+1$ dimensions
$$
U_{tt}U_{xx}-U_{tx}^2=-k\eqno(1)
$$
is the second order Monge-Amp\`ere equation. Here we will be interested in the case where $k$ is a constant. For $k=1$ we have the hyperbolic Monge-Amp\`ere equation which is equivalent [1] to the Born-Infeld equation [2]. The choice $k=-1$ yields the elliptic Monge-Amp\`ere equation that is related [3,4] to the equation for minimal surfaces [5]. Finally, $k=0$ corresponds to the homogeneous Monge-Amp\`ere equation that can be shown to be related to the Bateman equation [6]. The Born-Infeld, minimal surfaces and Bateman equations can be treated simultaneously as
$$
(k^2+\phi_x^2)\phi_{tt}-2\phi_x\phi_t\phi_{xt}+(k^2\alpha+\phi_t^2)\phi_{xx}=0\eqno(2)
$$
where
$$
\alpha\equiv k^2-k-1\eqno(3)
$$
and we should keep in mind the trivial identities $\alpha k^2=-k$ and
$\alpha k=-k^2$.

The Born-Infeld equation was introduced in 1934 as a nonlinear generalization of Maxwell's electrodynamics. It is the simplest wave equation in 1+1 dimensions, that preserves Lorentz invariance and is nonlinear.
This equation is integrable [7,8] and has a multi-Hamiltonian structure [9].
The Bateman equation was introduced in 1929 and is related with
hydrodynamics. This equation has a very interesting behavior [10]. If
$\phi(x,t)$ is a solution of (2), for $k=0$, so is any function of it
(covariance of (2)). Also, (2) can be derived from an infinite class
of inequivalent Lagrangian densities and is form invariant under
arbitrary linear transformations of the $(x,t)$ coordinates. The
equation for minimal surfaces gives the surface $z=\phi(x,t)$ in the three-dimensional space that spans a given contour and has the minimum area. This is the Plateau's problem and has interest both in physics and mathematics.

In this paper we will obtain Lax representations for $(1)$ and $(2)$
since both systems are related. A scalar dispersionless Lax
representation as well a matrix dispersive Lax representation will be
given. As far as the authors can say this is the first example of a
system where both Lax pairs are present. In fact our results suggest
that many other systems, which have both an infinite number of local and
nonlocal charges, are likely to have such characteristic.

This paper is organized as follows. In Section 2 we review the Bianchi
transformation which relates (1) and (2). This is the
Proposition 2.1 
that unifies the results obtained in [1,3,4]. With this transformation
we can easily translate results  from the system (1) to system
(2) and vice-versa. The existence of this Bianchi transformation is due to the fact
that both (1) and (2) can be rewritten in a hydrodynamic type equation
(polytropic gas). In Section 3, using results from [8,11,12], we obtain
the dispersionless Lax representation of (1) (Proposition 3.1) and
write the two sets of local conserved charges densities for the Monge-Amp\`ere
equation. In Section 4 we generalize the results of [5,13,14] concerning
the matrix Lax representation for minimal surfaces through its
correspondence with the sigma model. We obtain a matrix Lax
representation for a two parameter equation for minimal surfaces which
includes (2) for particular choices of the parameter (Proposition
4.4). From this Lax representation we give the nonlocal conserved
charges densities of the system. In Section
5 we  write explicitly the Lax representations, obtained in the
previous sections, for the Monge-Amp\`ere system (1) using the Bianchi
transformation (Proposition 5.1). Finally we present our conclusions in Section 6.
\bigskip
\noindent {\bf 2. {Bianchi Transformation}}
\medskip

In order to see the connection between (1) and (2) (see Equation (16))we have to express these equations in the form of equations of hydrodynamic type [15]. Following [9,16] we first introduce the potentials $a$ and $b$, defined as
$$
\eqalign{
a=&U_x\cr
b=&U_t
}\eqno(4)
$$
Then, Equation (1) can be expressed as a first order system
$$
\eqalign{
&k(a_t-b_x)=0\cr
&b_t={1\over a_x}(b_x^2-k)
}\eqno(5)
$$
which is the natural starting point for a Hamiltonian treatment of
Monge-Amp\`ere equations (1).

Now, introducing
$$
\eqalign{
u=&-{b_x\over a_x}\cr
v=&a_x
}\eqno(6)
$$
the Monge-Amp\`ere equation can be written in the following hydrodynamic type equation form
$$
\eqalign{
u_t+uu_x+kv^{-3}v_x=&0\cr
k(v_t+(uv)_x)=&0
}\eqno(7)
$$ 

Equation (2) follows from the Lagrangian
$$
{\cal L}=\sqrt{k^2+\phi_x^2+\alpha\phi_t^2}\eqno(8)
$$
We stress that the Bateman equation can be obtained from a large class of inequivalent Lagrangian. However, we will use this one and the limit $k\to 0$ will give us results for the Bateman equation.

Since (8) has no $\phi$ dependence (2) can be written as a conservation law given by
$$
\partial_x\left({\partial{\cal L}\over\partial\phi_x}\right)+
\partial_t\left({\partial{\cal L}\over\partial\phi_t}\right)=0\eqno(9)
$$
This result allows us to rewrite (2) as a set of coupled first order nonlinear equations. Following [9,16] let us express (2) as the integrability condition of a first-order system given by
$$
\eqalign{
\psi_x=&-{\partial{\cal L}\over\partial\phi_t}=-{\alpha\phi_t\over\sqrt{k^2+\phi_x^2+\alpha\phi_t^2}}\cr
\psi_t=&{\partial{\cal L}\over\partial\phi_x}={\phi_x\over\sqrt{k^2+\phi_x^2+\alpha\phi_t^2}}\cr
}\eqno(10)
$$
Introducing the variables
$$
\eqalign{
r=&\phi_x\cr
s=&\psi_x\cr
}\eqno(11)
$$
we get from (10)
$$
\eqalign{
\phi_t=&-\alpha s\sqrt{k^2+r^2\over 1-\alpha s^2}\cr
\psi_t=&r\sqrt{1-\alpha s^2\over k^2+r^2}
}\eqno(12)
$$
So, the one-forms
$$
\eqalign{
d\phi=&r \,dx-\alpha s\sqrt{k^2+r^2\over 1-\alpha s^2}\,dt\cr
d\psi=&s \,dx+r\sqrt{1-\alpha s^2\over k^2+r^2}\,dt\cr
}\eqno(13)
$$
are exact and its closure give us the equations
$$
\eqalign{
r_t=&-{\alpha rs\over\sqrt{(k^2+r^2)(1-\alpha s^2)}}\,\,r_x-\alpha\sqrt{k^2+r^2\over(1-\alpha s^2)^3}\, \,s_x\cr
s_t=&k^2\sqrt{1-\alpha s^2\over (k^2+r^2)^3}\,\,r_x-{\alpha rs\over\sqrt{(k^2+r^2)(1-\alpha s^2)}}\,\,s_x\cr
}\eqno(14)
$$
Now the amazing fact is that Equation (14) is also related with Equation (7) by a special transformation. For the case $k=1$ this transformation is known as the Verosky transformation [9]. We can easily check that the following $k$ generalized Verosky transformation
$$
\eqalign{
u=&{\alpha rs\over\sqrt{(k^2+r^2)(1-\alpha s^2)}}\cr
kv=-k&\sqrt{(k^2+r^2)(1-\alpha s^2)}\cr
}\eqno(15)
$$
links (14) with (7). From the diagram 
$$
\left.\eqalign{
{\rm Eq. (1)}\,\,\Rightarrow\,\,U\,\,{\buildrel{\rm Eq. (4)}\over\longrightarrow}\,\,\, a,b\,\,\,{\buildrel{\rm
Eq. (6)}\over\longrightarrow}\,\, u,v\,\,\Rightarrow&\,\,\,
U=U(u,v)\cr
\noalign{\vskip -5pt}%
\Downarrow\ \ \ \ \ \ &\cr
\noalign{\vskip -5pt}%
{\rm Eq. (7)}\ \ \ &\cr
\noalign{\vskip -5pt}%
\Uparrow\ \ \ \ \ \ &\cr
\noalign{\vskip -5pt}%
{\rm eq. (2)}\,\,\Rightarrow\,\,\phi\,\,{\buildrel{\rm Eq. (11)}\over\longrightarrow}\,\, r,s\,\,{\buildrel{\rm
Eq. (15)}\over\longrightarrow}\,\, u,v\,\,\Rightarrow&\cases{u=u(\phi)\cr v=v(\phi)}\cr
}
\right\}\Rightarrow U=U(\phi)
$$
we are led to  the  proposition [1,3,4]:
\medskip
\noindent
{ \bf Proposition 2.1}
{\it The Monge-Amp\`ere Equation $(1)$ and Equation $(2)$ are related by the
following Bianchi transformation
$$
\eqalign{
U_{tt}=&{k-\phi_t^2\over\sqrt{k^2+\phi_x^2+\alpha\phi^2_t}}\cr
U_{tx}=&{-\phi_x\phi_t\over\sqrt{k^2+\phi_x^2+\alpha\phi^2_t}}\cr
U_{xx}=&{-(k^2+\phi_x^2)\over\sqrt{k^2+\phi_x^2+\alpha\phi^2_t}}
}\eqno(16)
$$}
\bigskip
\noindent {\bf 3. {Dispersionless Lax Representation: Local Conserved Charges}}
\medskip

Equation (7) for $k=1$ corresponds to the equations of isentropic, polytropic gas dynamics with the adiabatic index $\gamma=-1$ [9]. This system is known as a Chaplygin gas [17]. For $k=0$ (7) is the Riemann equation [11] and in this case the transformation (15) give us $ u=-{\phi_t\over\phi_x}$.

In [12] the polytropic gas dynamics [18] equations
$$
\eqalign{
u_t+uu_x+v^{\gamma-2}v_x=&0\,\,,\quad \gamma\ge2\cr
v_t+(uv)_x=&0
}\eqno(17)
$$
were derived from the following dispersionless nonstandard Lax representation
$$
\eqalign{
L=&p^{\gamma-1}+u+{v^{\gamma-1}\over(\gamma-1)^2}p^{-(\gamma-1)}\,\,,\quad \gamma\ge2\cr\cr
{\partial L\over\partial t}=&{(\gamma-1)\over\gamma}
\left\{\left(L^{\gamma\over\gamma-1}\right)_{\ge1},L\right\}
}\eqno(18)
$$
Here $\{A,B\}={\partial A\over\partial x}{\partial B\over\partial p}-
{\partial B\over\partial x}{\partial A\over\partial p}$ and 
$\left(L^{\gamma\over\gamma-1}\right)_{\ge1}$ stands for the purely nonnegative (without $p^0$ terms) part of the polynomial in $p$. In (18) $L^{1\over\gamma-1}$ was expanded around $p=\infty$. A Lax description for the Chaplygin gas like equations
$$
\eqalign{
u_t+uu_x+{v_x\over v^{\beta+2}}=&0\,\,,\quad \beta\ge1\cr
v_t+(uv)_x=&0
}\eqno(19)
$$
was obtained in [8] in connection with the Born-Infeld equation and it is given by
$$
L=p^{-(\beta+1)}+u+{v^{-(\beta+1)}\over(\beta+1)^2}\,p^{\beta+1}\,\,,\quad \beta\ge1\eqno(20)
$$
with
$$
{\partial L\over\partial t}={(\beta+1)\over\beta}
\left\{\left(L^{\beta\over\beta+1}\right)_{\le1},L\right\}
\eqno(21)
$$
where $L^{1\over\beta+1}$ is expanded around $p=0$.

In view of these results we have the proposition:
\medskip
\noindent
{ \bf Proposition 3.1}
{\it For $\beta=1$, the Lax operator
$$
L=p^{-2}+u+{k\over 4v^2}p^2\eqno(22)
$$
where
$$
\left(L^{1/2}\right)_{\le1}=p^{-1}+{1\over2}up
$$
reproduces $(7)$. In terms of the variables $a$ and $b$ the Lax representation $(22)$ assumes the form
$$
\eqalign{
L=&p^{-2}-{b_x\over a_x}+{k\over4 a_x^2}p^2\cr
{\partial L\over\partial t}=&2
\left\{\left(L^{1/2}\right)_{\ge1},L\right\}
}\eqno(23)
$$
and yields the Monge-Amp\`ere equation as expressed in $(5)$.}
\medskip

This proposition is the first main result of our paper. This is a 
dispersionless Lax representation, a dispersive one will be obtained
in Section 5 (see Proposition 5.1).

Conserved charges for the Chaplygin gas like equations (19) can be easily obtained from (20) through [8,12]
$$
{\cal H}_n=\hbox{Tr}\,L^{n+{\beta+2\over\beta+1}},\,\,n=0,1,2,3,\dots\eqno(24)
$$
This conserved charges were obtained by expanding $L^{1\over\beta+1}$ around
$p=0$. An alternate expansion around $p=\infty$ is possible and it gives us a second set of conserved charges through
$$
\widetilde{\cal H}_n=\hbox{Tr}\,L^{n-{1\over\beta+1}},\,\,n=0,1,2,3,
\dots\eqno(25)
$$
Both set of densities for (24) and (25) can be expressed in closed form [8]. They are
$$
\eqalign{
H_n=&(n+1)!\hbox{C}^{(n+1)(\beta+1)+1\over(\beta+1)}_{n+1}
\sum_{m=0}^{\left[{n+1\over2}\right]}
\left(-\prod_{\ell=0}^m{-1\over \ell(\beta+1)+1}\right)
{u^{n-2m+1}\over m!(n-2m+1)!}{v^{-m(\beta+1)}\over(-\beta-1)^m}\cr
{\widetilde H}_n=&
n!(-\beta-1)^{2\over\beta+1}
\hbox{C}^{n(\beta+1)-1\over(\beta+1)}_{n}
\sum_{m=0}^{\left[{n\over2}\right]}
\left(\prod_{\ell=0}^m{-1\over \ell(\beta+1)-1}\right)
{u^{n-2m}\over m!(n-2m)!}{v^{-m(\beta+1)+1}\over(-\beta-1)^m}\cr
}\eqno(26)
$$
The first densities $H_n$ for the Monge-Amp\`ere are
$$
\eqalign{
H_0=&-{3\over2}{b_x\over a_x}\cr
H_1=& {5\over 8}{1\over a_x^2}\left(3b_x^2+k\right)\cr
H_2=&-{35\over 16}{b_x\over a_x^3}\left(b^2_x+k\right)\cr
H_3=&{63\over 128}{1\over a_x^4}\left(5b_x^4+10b_x^2k+k^2\right)\cr
&\vdots\cr
}\eqno(27)
$$
and the first densities ${\widetilde H}_n$ are
$$
\eqalign{
{\widetilde H_0}=&-2k^2a_x\cr
{\widetilde H_1}=& -k^2b_x\cr
{\widetilde H_2}=&-{3\over 4}k^2{1\over a_x}\left(b_x^2+k\right)\cr
{\widetilde H_3}=& {5\over 8}k^2{b_x\over a_x^2}\left(b_x^3+3k\right)\cr
&\vdots\cr
}\eqno(28)
$$
\bigskip
\noindent {\bf 4. {Minimal Surfaces and Sigma Models}}
\medskip
 In this section we will generalize some results of [5,13,14] where a
matrix Lax representation for the minimal surface equation (Eq. (2)
with $k=-1$) was obtained.

Let $g$ be a $2\times 2$ matrix function with components
$$
g_{11}= {k_1 +a^2\over \omega}\, , \quad g_{12}=g_{21}={ab\over \omega}\quad
\hbox {and}\quad g_{22}={ k_2 +b^2\over \omega}\, ,
$$
where $k_1$ and $k_2$ are arbitrary constants, not vanishing simultaneously 
and $$
\varepsilon\omega^2=k_1 k_2 + k_1b^2 + k_2a^2\, , \quad \hbox{where}\quad 
\varepsilon=\pm 1. $$
Thus, $\det g=\varepsilon$. Note that $\varepsilon$ is not fully independent of $k_1$ and $k_2$.
$\omega^2>0$ when we are dealing with real fields. In the case
of complex fields $\varepsilon$ is independent of $k_1$ and $k_2$ . 

The sigma model equation can be written as 
$$ 
\partial_\alpha (g^{\alpha\beta}g^{-1}\partial_\beta g)=0\, ,
\eqno(29)
$$
where $g^{\alpha\beta}$ are the components of $g^{-1}$. As shown in
[5] and [13] the Lax representation of (29) is
$$
\varepsilon^{\alpha\beta}\partial_\beta\psi={1\over \lambda^2
+\varepsilon}\left[\lambda g^{\alpha\beta}-
\varepsilon\varepsilon^{\alpha\beta}\right](g^{-1}\partial_\beta g)\psi,
\eqno (30) 
$$ 
where $\varepsilon^{\alpha\beta}$ is  Levi-Civita tensor with 
$\varepsilon^{12}=1$, $\lambda$ is the spectral parameter, $\det
g=\varepsilon$ and $\varepsilon=\pm1$.

\medskip

Now, let us see how a Lax representation for (2) can be obtained from
(30). First, let $M_3$ be a 3-dimensional manifold with metric 
$$
ds^2|_{M_3}=k_1dt^2 +  k_2dx^2 + dz^2 \quad (k_1\ne 0,\, \, k_2\ne 0)
$$
and $z=\phi(t,x)$ define a graph of a regular surface $S$ in $M_3$.
The induced metric on $S$ is given by
$$ 
ds^2|_S=(k_1+\phi_t^2)dt^2 + (k_2 + \phi_x^2)dx^2 + 2\phi_x\phi_tdxdt.
$$
If $a=\phi_t$ and $b=\phi_x$, then $g$ is a metric tensor on $S$.
Surface $S$ is called minimal if its mean curvature $H$
vanishes. Minimality condition leads to the equation
$$
g^{\alpha\beta}\partial_\alpha\partial_\beta\phi=0,
$$  
or
$$
(k_1+\phi^2_t)\phi_{xx} - 2\phi_x\phi_t\phi_{xt}+(k_2+\phi_x^2)\phi_{tt}=0\, . \eqno(31)
$$
There is a parametrization of the minimal surfaces where the
minimality condition reduces to the Laplace equation in 2-dimensions.
Let $X:S\rightarrow M_3$ define a parametrization of $S$ in
$M_3$. This parametrization is called isothermal [19,20],
if
$$
\langle X_u\, X_u\rangle=\varepsilon\langle X_v\, X_v\rangle \eqno (32)
$$
$$
\langle X_u\, X_v\rangle=0 \quad \quad (\varepsilon=\pm 1)   \eqno (33)
$$

\noindent
{\bf Proposition 4.1.}
{ \it $S$ is a minimal surface if and only if $X_{uu}+\varepsilon
X_{vv}=0$, where $X$ is an isothermal parametrization.}

A connection between the above two different parametrizations may be
obtained from the following two propositions:

\noindent
{\bf Proposition 4.2.}
{\it Let $z=\phi(t,x)$ define  a regular surface $S$. Parametrization $X:S\rightarrow M_3$ is 
isothermal if and only if the following equations are satisfied
$$
\eqalign{
(k_1 +\phi_t^2)t_u &= -\omega x_v - \phi_t\phi_x x_u\,, \cr
(k_2+ \phi_t^2)t_v &= -\omega x_u - \phi_t\phi_x  x_v \, .\cr}\eqno (34) 
$$ }
The proof of the  Proposition 4.2 can be done in the  following way.
The Equation (33) can be written as 
$$ 
x_u(k_2x_v + \phi^2_xx_v + \phi_t\phi_xt_v)+t_u(k_1t_v  + \phi_t\phi_xx_v + 
\phi^2_tt_v)=0 $$
and it is equivalent to the system
$$
\cases{
t_u = \lambda^{-1}[(k_2 +\phi_x^2)x_v + \phi_t\phi_xt_v] & \cr
\noalign{\vskip 7pt}
x_u = \lambda^{-1}[(k_1 + \phi_t^2)t_v +\phi_t\phi_x x_v]\, . & \cr }
$$
Inserting expressions for $t_u$ and $x_u$ into (32), it can be found that 
$$
\varepsilon {\lambda}^2=(k_1k_2 + k_1\phi_x^2 + k_2\phi_t^2).
$$
Hence, $\lambda =\omega$ and
$$
\cases{
t_u =\omega^{-1}[(k_2 +\phi_x^2)x_v + \phi_t\phi_xt_v] & \cr
\noalign{\vskip 7pt}
 x_u =\omega^{-1}[(k_1 + \phi_t^2)t_v +\phi_t\phi_x x_v ]. & \cr }
$$
That is equivalent to (34).

\medskip

Proposition 4.1 and  4.2 imply next proposition:

\medskip

\noindent 
{\bf Proposition 4.3.}
{\it Let $x$ and $t$ be harmonic functions of $u$ and $v$. 
Let a differentiable function  $\phi(t,x)$ be defined by $(34)$. Then the 
function 
$\phi(t,x)$ is a harmonic  function  of $u$ and $v$ if and only if it satisfies the
minimality condition $(31)$.} 

\medskip

Let us consider Equation (31),
where  $k_1$ and $k_2$ are arbitrary constants. We have four distinct cases:

\noindent \item{(i)}{$k_1k_2>0$}. 

\itemitem{(1)\ }{$k_1>0, k_2>0$. This is equivalent to the equation of minimal surface in 
${\bf R^3 }$ or elliptic Monge-Amp\`ere equation ($k_1=k_2=-k=1$).}

\itemitem{(2)\ }{ $k_1>0, k_2<0$. This is equivalent to the equation of minimal surface in ${\bf M_3}$ (3-dimensional Minkowski space with metric $(1,1,-1)$).}

\item{(ii)}{$k_1k_2<0$. This  is equivalent to the Born-Infeld
equation (which is the equation of a minimal surface in 
 a 3-dimensional Minkowski space with metric $(-1,1,1)$) or hyperbolic
 Monge-Amp\`ere equation ($-k_1=k_2=k=1$)}.

\noindent We have the following cases which do not arise from the
 embedding problem  in $M_3$:

\item{(iii)}{$k_1k_2=0$, but not simultaneously vanishing. This is a 
new type of equation.} 

\item{(iv)}{$k_1=k_2=0$. This is  Bateman equation or homogeneous Monge-Amp\`ere equation ($k_1=k_2=k=0$).} 

The next proposition is very important since it provides the Lax pair
for systems that include Equation (2): 

\noindent
{\bf Proposition 4.4}
{\it  Let $\phi$ be a differential function of $t,x$  and let $a=\phi_t,\, b=\phi_x$.
Then Equation $(31)$
solves the sigma model Equation $(29)$, if $k_1,k_2$ not vanish simultaneously.

 \noindent If  $k_1=k_2=0$  Equation $(31)$ solves the sigma model
 Equation  $(29)$ for another matrix $g$, namely,
$$   
g_{11}=a_1{\phi_t\over \phi_x},\quad g_{12}=g_{21}=b_1 \quad \hbox{and}\quad  
g_{22}=a_2{\phi_x\over \phi_t}, $$
where $a_1,a_2,b_1$ are constants. The Lax pair of $(31)$ is then given
by $(30)$.
}

In the next section we will use the last Proposition to obtain the Lax
representations for the Monge-Amp\`ere equations (1). In doing so we
will return to our original parameter $k$ instead of working with the
parameters $k_1$ and $k_2$. It is just a matter of scale
transformation either in formula for $ds^2|_{M_3}$ or in Equation (31)
(redefining $x$ and $t$) to give $k_1=\pm1$ and $k_2=\pm1$. Also,
we will set $\varepsilon=1$ in the next section.
\bigskip
\noindent {\bf 5. {Matrix Lax Representation: Nonlocal Conserved Charges}}
\medskip

Now we can write the Lax pairs for (1). First, let us give the Lax pairs
for (2) more explicitly. Equation (30) can be rewritten in the form
$$
\eqalign{
{\partial \psi\over\partial x}={1\over
\lambda^2+1}&\left[\lambda\left(g^{11}A+g^{12}B\right)-B\right]\psi\cr
{\partial \psi\over\partial t}=-{1\over
\lambda^2+1}&\left[\lambda\left(g^{21}A+g^{22}B\right)+A\right]\psi\cr
}\eqno(35)
$$ 
where
$$
A=g^{-1}\partial_t g\,,\quad B=g^{-1}\partial_x g\eqno(36)
$$
From (36) it follows the identity
$$
{\partial A\over\partial x}-{\partial B\over\partial t}-[A,B]=0\eqno(37)
$$
The integrability of (35) yields the equations
$$
\displaylines{
\hfill\det g= 1\hfill(38)\cr
\hfill(g^{11}A-g^{12}B)_t+(g^{21}A+g^{22}B)_x=0 \hfill(39)\cr
}
$$
From the Proposition 4.4 we have for $k\ne 0$
$$
g={1\over\sqrt{-k(1+\phi_x^2)+\phi_t^2}}\pmatrix{-k+\phi_t^2&\phi_t\phi_x\cr
\noalign{\vskip 5pt}%
\phi_t\phi_x &1+\phi_x^2}\eqno(40)
$$
and for $k=0$ (setting $a_1=a_2=\sqrt{2}$ and $b_1=1$)
$$
g=\pmatrix{\sqrt{2}\,\displaystyle{\phi_t\over\phi_x}&1\cr
\noalign{\vskip 5pt}%
1&\sqrt{2}\,\displaystyle{\phi_x\over\phi_t}}\eqno(41)
$$
With this choice (37) and (38) are trivial identities and (39) is
identical to Equation (2), i.e., to the minimal surface equation for
$k=-1$, Born-Infeld equation for $k=1$ and Bateman equation for $k=0$.

The Bianchi transformation (16) for $k\ne 0$ assumes the form
$$
\eqalign{
\sqrt{-k}\,U_{tt}=&{-k+\phi^2_t\over\sqrt{-k(1+\phi_x^2)+\phi_t^2}}\cr
\sqrt{-k}\,U_{tx}=&{\phi_x\phi_t\over\sqrt{-k(1+\phi_x^2)+\phi_t^2}}\cr
\sqrt{-k}\,U_{xx}=&{1+\phi^2_x\over\sqrt{-k(1+\phi_x^2)+\phi_t^2}}\cr
}\eqno(42)
$$
and (40) in terms of $U$ can be written as
$$
g=\sqrt{-k}\pmatrix{U_{tt}&U_{tx}\cr
\noalign{\vskip 5pt}%
U_{tx}& U_{xx}}\eqno(43)
$$
In this way (35) with (43) give us the matrix Lax representation for
the hyperbolic Monge-Amp\`ere equation ($k=1$) and elliptic
Monge-Amp\`ere equation ($k=-1$). Let us observe that (1) for $k\ne0$
follows from (38) while Equations (37) and (39) are trivial
identities. We can also express (43) in terms of variables $a$ and $b$
defined in (4) by
$$
g=\sqrt{-k}\pmatrix{b_{t}&b_{x}\cr
\noalign{\vskip 5pt}%
b_{x}& a_{x}}\eqno(44)
$$
and (5) follows easily since $a_t=b_x$ is a trivial identity and $\displaystyle\det
g=-k(b_ta_x-b_x^2)=1$. The Bianchi transformation (16) for $k=0$
yields
$$
{\phi_t\over\phi_x}={U_{tt}\over U_{tx}}\,,\quad{\phi_x\over\phi_t}={U_{xx}\over U_{tx}}\eqno(45)
$$
and (41) in terms of $U$ can be written as
$$
g=\pmatrix{\sqrt{2}\,\displaystyle{U_{tt}\over U_{tx}} &1 \cr
\noalign{\vskip 5pt}%
1&\sqrt{2}\,\displaystyle{U_{xx}\over U_{tx}} }\eqno(46)
$$
and $\det g=1$ reproduces (1) for $k=0$. In terms of the variables $a$
and $b$ we have
$$
g=\pmatrix{\sqrt{2}\,\displaystyle{b_{t}\over b_{x}} &1 \cr
\noalign{\vskip 5pt}%
1&\sqrt{2}\,\displaystyle{a_{x}\over b_{x}} }\eqno(47)
$$
which give us (5) for $k=0$.

So, we have the following proposition:

\medskip
\noindent
{ \bf Proposition 5.1}
{\it The Lax pair $(35)$ with $(44)$ or $(47)$ yields the Monge-Amp\`ere
equations as expressed in $(5)$ for $k\ne0$ and  $k=0$,  respectively.}
\medskip

This proposition is the second main result of our paper. This is a 
matrix dispersive Lax representation. 

In Section 3, using the dispersionless Lax representation for the
Monge-Amp\`ere equations (1), we were able to derive two sets of
infinite number of local conserved charges. Now, using (35) it will
possible to find infinitely non local conserved ones. Let us denote
$M=-(g^{11}A+g^{12}B)$ and $N=g^{21}A+g^{22}B$, then the Lax pair (35)
can be written as
$$
\eqalign{
(\lambda^2+1)\psi_x=& -\lambda M\psi- g^{-1}g_x\psi\cr
(\lambda^2+1)\psi_t=& -\lambda N\psi- g^{-1}g_t\psi\cr
}
$$
or
$$
\eqalign{
(g\psi)_x=&-\lambda g M\psi-\lambda^2 g\,\psi_x\cr
(g\psi)_t=&-\lambda g N\psi-\lambda^2 g\,\psi_t\cr
}\eqno(48)
$$
Let us assume that function $\psi$ is analytical in the parameter
$\lambda$ and can be expanded as
$$
\psi=\psi_0+\lambda\psi_1+\lambda^2\psi_2+\cdots\eqno(49)
$$
Then, Equations (48) imply
$$
\eqalign{
\psi_0=&g^{-1}\cr
(g\psi_1)_x=&-gMg^{-1}\cr
(g\psi_1)_t=&-gNg^{-1}\cr
(g\psi_2)_x=&g_xg^{-1}+gMg^{-1}\partial_x^{-1}(gMg^{-1})\cr
(g\psi_2)_t=&g_tg^{-1}+gNg^{-1}\partial_t^{-1}(gNg^{-1})\cr
\vdots&\cr
}\eqno(50)
$$
and we have now infinitely many conserved laws in the form
$(X_n)_x=(T_n)_t$ where the densities are
$$
\eqalign{
X_1=&N\cr
T_1=&M\cr
X_2=&g^{-1}g_t+(\partial_t^{-1}N)N\cr
T_2=&g^{-1}g_x+(\partial_x^{-1}M)M\cr
\vdots&\cr
}\eqno(51)
$$
Now we can use (44) and (47) to express the densities $T_n$ in terms
of variables $a$ and $b$. 
\bigskip\vfill\eject
\noindent {\bf 6. {Conclusion}}
\medskip

In this paper we have obtained the Lax representation of the
Monge-Amp\`ere equations (1). In Section 2 the Bianchi transformation
relating equations (1) and (2) was given (Proposition 2.1). This transformation
allowed us to translate results obtained for one equation to the
other. In Section 3
the dispersionless Lax pair for (1) as well the local conserved
densities were given (Proposition 3.1). In Section 4 the
correspondence between sigma models and a two parameter equation for minimal
surfaces was given and the matrix Lax pair for equation (2) 
 was obtained (Proposition 4.4). A Lax representation for the
system (1) as well the nonlocal conserved densities were given in
Section 5 (Proposition 5.1).

The algebra of the local and nonlocal
charges that follows from (27), (28) and (51) as well the multiHamiltonian
formulation of the Monge-Amp\`ere equations (1) will be the
subject of a future publication. Some results on this line for the
second order homogeneous Monge-Amp\`ere equation were already obtained
in [21,22]. As we have pointed, the homogeneous Monge-Amp\`ere equation
has an infinite number of inequivalent Lagrangians and somehow this
should be reflected in its Lax representation. This also deserves
further clarifications.
\bigskip
\leftline{\bf Acknowledgments}
\medskip
 
This work was partially  supported by 
 the Scientific and Technical Research Council of Turkey 
(T\"UB\.ITAK), Turkish Academy of Sciences (T\"UBA) and
 CNPq, Brazil.

\vskip 2.0truecm

\vfill\eject

\leftline{\bf References}
\bigskip

\item{1.} O.I. Mokhov and Y. Nutku, Lett. Math. Phys. {\bf 32}, 121
 (1994).

\item {2.} M. Born and L. Infeld, Proc. R. Soc. London {\bf A144}, 425 (1934).

\item{3.} K. J\"orgens, Math. Annal. {\bf 127}, 130 (1954).

\item{4.} E. Heinz, Nach. Akad. Wissensch. in G\"ottingen
Mathem.-Phys. Klasse {\bf IIa}, 51 (1952).

\item{5.} M. G\"urses, Lett. Math. Phys. {\bf 44}, 1 (1998).

\item{6.} H. Bateman, Proc. R. Soc. {\bf A125}, 598 (1929).

\item{7.} B. M. Barbishov and N. A. Chernikov, Sov. Phys. JETP {\bf 24}, 93 (1966).

\item{8.} J. C. Brunelli and A. Das,  Phys. Lett. {\bf B426}, 57 (1998).

\item{9.} M. Arik, F. Neyzi, Y. Nutku, P. J. Olver and J. Verosky, J. Math. Phys. {\bf 30}, 1338 (1988).

\item{10.} D. B. Fairlie, J. Govaerts and A. Morozov, Nucl. Phys. {\bf
B373}, 214 (1992).

\item{11.} J. C. Brunelli, Rev. Math. Phys. {\bf 8}, 1041 (1996).

\item{12.} J. C. Brunelli and A. Das, Phys. Lett. {\bf A235}, 597 (1997).

\item{13.} M. G\"urses and A. Karasu, Internat. J. Modern Phys. A. {\bf 6}, (1991)

\item{14.} M. G\"urses, Lett. Math. Phys. {\bf 26}, (1992).

\item{15.} B. Dubrovin and S. Novikov, Russ. Math. Surv. {\bf 44}, 35 (1989).

\item{16.} Y. Nutku, J. Math. Phys. {\bf 26}, 1237 (1985).

\item{17.} K. P. Stanyukovich, ``Unsteady Motion of Continuous Media'' (Pergamon, New York, 1960), p. 137.

\item{18.} P. J. Olver and Y. Nutku, J. Math. Phys. {\bf 29}, 1610 (1988).

\item{19.} M. do Carmo, ``Differential Geometry of Curves and Surfaces'' (Prentice-Hall, New Jersey, 1976).

\item{20.} U. Dierken, S. Hildebrandt, A. K\"unster and O. Wohlrab, ``Minimal Surfaces I'', Grundlehren der Mathematishen Wissenschaften, No. 295 (Springer-Verlag, Berlin-Heidelberg, 1992).

\item{21.} Y. Nutku and \"O. Sario{\v g}lu, Phys. Lett. {\bf A173}, 270 (1993).

\item{22.} Y. Nutku, J. Phys. {\bf 29A}, 3257 (1996) 

\end